\documentclass[onecolumn,preprintnumbers,amsmath,amssymb]{revtex4}
\usepackage{amssymb}
\usepackage{txfonts}
\usepackage{amsmath}

\usepackage{graphicx}
\usepackage{dcolumn}
\usepackage{bm}

\begin{document}

\title{Quasinormal modes of Dirac field perturbation in Reissner-Nordstr\"{o}m black hole surrounded by quintessence}
\author{Wang Chun-Yan$^1$}
\email{wangchunyan5800@yahoo.cn}
\author{Zhang Yu$^1$}
\author{Gui Yuan-Xing$^1$}
\email{guiyx@dlut.edu.cn}
\author{Lu Jian-Bo$^1$}
 \affiliation{1.School of Physics and Optoelectronic Technology, Dalian
University, Dalian 116024, China\\}

\begin{abstract}
The dirac quasinormal modes of the Reissner-Nordstr\"{o}m black hole
surrounded by quintessence are investigated using the third WKB
approximation. We find that the magnitude of the imaginary part of
the quasinormal mode frequencies increases firstly and then
decreases as the charge $Q$ increases, but it decreases as the
absolute value of  $w_{q}$ increases. The magnitude of the imaginary
part of quasinormal complex frequencies is smaller than those with
no quintessence. That is to say, the dirac field damps more slowly
due to the presence of quintessence.
\end{abstract}


\maketitle

\section {Introduction}
The concept of QNMs (quasinormal
modes)\cite{Kokkotas01,Nollert,Chandrasekhar,Hod,Cardoso01,Cardoso02,Berti01,Berti02,Konoplya01,Konoplya02,Abdalla,Lixz01,Wangb01,Wangb02,Zhanghb,Ortega01,Ortega02}was
firstly pointed out by Vishveshwara\cite{Vishveshwara} in
calculations of the scattering of gravitational waves by a black
hole. Due to the waves emitted by perturbed black holes there are no
normal mode oscillations but instead of quasi-normal mode
frequencies, with the real part representing the actual frequency of
the oscillation and the imaginary part representing the damping.
Quasinormal mode frequencies are the solutions of the perturbation
equations, which satisfy the boundary conditions appropriate for
purely ingoing waves at the horizon and purely outgoing waves at
infinity. Quasinormal modes are regarded as the "characteristic
sound" of black holes. Indeed, quasinormal ringing could provide the
direct evidence of the existence of black holes if observed by LIGO
(the Laser Interferometric Gravitational Wave Observatory), GEO600,
VIRGO, TAMA300, etc.\\
Recently, Kiselev\cite{Kiselev} considered Einstein's field
equations for a black hole surrounded by the quintessential matter
and obtained a new solution, which depend on the state parameter
$w_{q}$ of the quintessence. From then on, several people
investigate the property of the black hole surrounded by
quitessence\cite{Chensb01,Chensb02,Zhang01,Zhang02,Chenjh01,Chenjh02,Lixz02,Varghese,Liuml},
and get some valuable conclusion. In this paper, we use the
third-order WKB approximation to explore the quasinormal modes of
massless Dirac
perturbation\cite{Cho,Giammatteo,Jing01,Jing02,Jing03} in
RN(Reissner-Nordstr\"{o}m) black hole surrounded by quintessence.

\section {Dirac equation in the spacetime of the Reissner-Nordstr\"{o}m black hole
surrounded by quintessence}

The metric for the RN black hole surrounded by the static
spherically-symmetric quintessence can be expressed
as\cite{Kiselev,Varghese}
 \begin{equation}\label{eq:1}
 ds^{2}=-fdt^{2}+f^{-1}dr^{2}+r^{2}(d\theta^{2}+sin^{2}\theta
d\phi^{2}),\end{equation}
 with
 \begin{eqnarray}
 f=1-\frac{2M}{r}+\frac{Q^{2}}{r^2}-\frac{c}{r^{3w_{q}+1}}
 \end{eqnarray}
where $M$ represents the black hole mass, $w_{q}$ is the
quintessential state parameter, $c$ is the normalization factor
related to $\rho_{q}=-\frac{c}{2}\frac{3w_{q}}{r^{3(1+w_{q})}}$, and
$\rho_{q}$ is the density of quintessence.
\\ The massless Dirac equation in a general background spacetime can be
written as\cite{Brill}
 \begin{eqnarray}
  [\gamma^a e_a^\mu(\partial _\mu+\Gamma_\mu)]\Psi=0, \label{Di}
\end{eqnarray}
where $\gamma^a$ are the Dirac matrices,
\begin{eqnarray}
\gamma^{0}= \left(
\begin{array}{cc}
-i&0\\0&i
\end{array}\right),\ \ \
\gamma^{i}=\left(
\begin{array}{cc}
0&-i\sigma^{i}\\i\sigma^{i}&0
\end{array}\right),\ i=1,2,3,
\end{eqnarray}
$e_a^\mu$ is the inverse of the tetrad $e_\mu^a$, which defined by
the metric $g_{\mu\nu}$,
\begin{eqnarray}
g_{\mu\nu}=\eta_{ab}{e_{\mu}}^{a}{e_{\nu}}^{b}
\end{eqnarray}
with $\eta_{ab}={\rm diag}(-1,1,1,1)$ being the Minkowski metric.
\\$\Gamma_\mu $ is the spin
connection which is defined as
\begin{eqnarray}
 \Gamma_\mu=
\frac{1}{8}[\gamma^a,\gamma^b] e_a^\nu e_{b\nu;\mu}
\end{eqnarray}
where
$e_{b\nu;\mu}=\partial_{\mu}e_{b\nu}-\Gamma^{\alpha}_{\mu\nu}e_{b\alpha}$
is the covariant derivative of $e_{b\nu}$, and
$\Gamma^{\alpha}_{\mu\nu}$ is the Christoffel symbol.
\\we take the tetrad as
\begin{eqnarray}
    e_\mu^a=diag(\sqrt{f}, \frac{1}{\sqrt{f}}, r, r \sin \theta).
\end{eqnarray}
Then, the Dirac equation (\ref{Di}) becomes
\begin{eqnarray}
    \frac{\gamma^{0}}{\sqrt{f}}\frac{\partial \Psi}{\partial t}+\sqrt{f}
    \gamma^{1} \left(\frac{\partial }{\partial r}+\frac{1}{r}+\frac{1}{4 f}
    \frac{d f}{d r} \right) \Psi+\frac{\gamma^{2}}{r}(\frac{\partial }
    {\partial \theta}+\frac{1}{2}cot\theta)\Psi+\frac{\gamma^{3}}{r
    \sin\theta}\frac{\partial \Psi}{\partial \varphi}=0. \label{Di1}
\end{eqnarray}
Defining
\begin{eqnarray}
    \Psi=f^{-\frac{1}{4}}\Phi,
\end{eqnarray}
Eq. (\ref{Di1}) becomes
\begin{eqnarray}\label{Di2}
    \frac{\gamma^{0}}{\sqrt{f}}\frac{\partial \Phi}{\partial t}+\sqrt{f}
    \gamma^{1}\left(\frac{\partial }{\partial r}+\frac{1}{r} \right)
    \Phi+\frac{\gamma^{2}}{r}(\frac{\partial }{\partial \theta}+
    \frac{1}{2}cot\theta)\Phi+\frac{\gamma^{3}}{r \sin\theta}
    \frac{\partial \Phi}{\partial \varphi}=0. \label{Di2}
\end{eqnarray}
Introducing a tortoise coordinate change
\begin{eqnarray}
     r_{*}=\int\frac{dr}{f}
\end{eqnarray}
and the ansatz
\begin{eqnarray}
    \Phi=\left(
\begin{array}{c}
\frac{i G^{(\pm)}(r)}{r}\phi^{\pm}_{jm}(\theta, \varphi) \\
\frac{F^{(\pm)}(r)}{r}\phi^{\mp}_{jm}(\theta, \varphi)
\end{array}\right)e^{-i \omega t},
\end{eqnarray}
with spinor angular harmonics
\begin{eqnarray}
    \phi^{+}_{jm}=\left(
\begin{array}{c}
\sqrt{\frac{l+\frac{1}{2}+m}{2 l+1}}Y^{m-1/2}_l \\
\sqrt{\frac{l+\frac{1}{2}-m}{2 l+1}}Y^{m+1/2}_l
\end{array}\right), \ \ \ \ \ \ \ \ \ \ \ \  (for \ \ j=l+\frac{1}{2}),
\nonumber
\end{eqnarray}
\begin{eqnarray}
    \phi^{-}_{jm}=\left(
\begin{array}{c}
\sqrt{\frac{l+\frac{1}{2}-m}{2 l-1}}Y^{m-1/2}_l \\
-\sqrt{\frac{l+\frac{1}{2}+m}{2 l-1}}Y^{m+1/2}_l
\end{array}\right), \ \ \ \ \ \ (for \ \ j=l-\frac{1}{2}), \nonumber
\end{eqnarray}
Eq.(\ref{Di2}) can be written in the form
\begin{eqnarray}
\left(\begin{array}{cc}0  & -\omega \\ \omega  & 0
\end{array}\right) \left(
\begin{array}{c}F^{\pm} \\G^{\pm}\end{array}\right)-\frac{\partial}
{\partial r_*}\left(\begin{array}{c}F^{\pm}
\\G^{\pm}\end{array}\right)+\sqrt{f}\left(\begin{array}{cc}
\frac{k_{\pm}}{r}  & 0 \\ 0 &  -\frac{k_{\pm}}{r} \end{array}\right)
\left(
\begin{array}{c}F^{\pm} \\G^{\pm}\end{array}\right)=0.
\end{eqnarray}
The cases for $(+)$ and $(-)$ in the functions $F^{\pm}$ and
$G^{\pm}$ can be put together, and then the two decoupled equations
can be expressed in the form
\begin{eqnarray}
    \frac{d^2 F}{d r_*^2}+(\omega^2-V_1)F&=&0, \label{even}\\
    \frac{d^2 G}{d r_*^2}+(\omega^2-V_2)G&=&0, \label{odd}
\end{eqnarray}
with
\begin{eqnarray}
V_1&=&\frac{\sqrt{f}|k|}{r^2}\left(|k|\sqrt{f}+\frac{r}{2}\frac{d f}
{d r}-f\right), \ \ \ \left(for \ \  k=j+\frac{1}{2},\ \ \ \ \ \  \
\ \ \  and \ \ j=l+\frac{1}{2}\right), \label{V1} \\
V_2&=&\frac{\sqrt{f}|k|}{r^2}\left(|k|\sqrt{f}-\frac{r}{2}\frac{d f}
{d r}+f\right), \ \ \ \left(for \ \  k=-\left(j+\frac{1}{2}\right),\
\  and \ \ j=l-\frac{1}{2}\right). \label{V2}
\end{eqnarray}
The potentials $V_1$ and $V_2$ are supersymmetric partners derived
from the same superpotential. It has been demonstrated that
potentials related in this way have the same quasinormal mode
complex frequencies\cite{Anderson}. We shall concentrate just on Eq.
\ref{even} with potential $V_1$ in evaluating the quasinormal mode
frequencies for the massless Dirac field by the third WKB
approximation.

\section {WKB approximation and Massless Dirac quasinormal mode frequencies in the Reissner-Nordstr\"{o}m black hole surrounded by quintessence}
\begin{figure}
    \includegraphics[angle=0, width=0.6\textwidth]{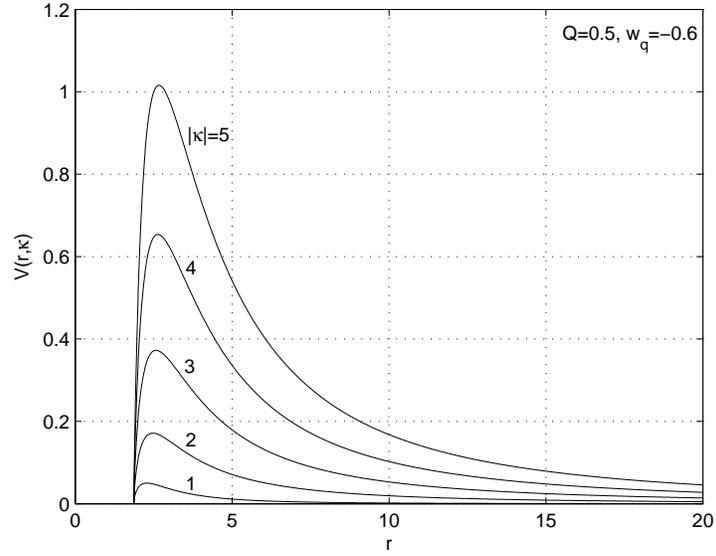}
\caption{Variation of the effective potential for the massless Dirac
field with $|\kappa|$ in RN black hole surrounded by quintessence
for $w_{q}=-0.6$, $Q=0.5$ and $c=0.001$.\label{fig:fig1}}
\end{figure}

\begin{figure}
    \includegraphics[angle=0, width=0.6\textwidth]{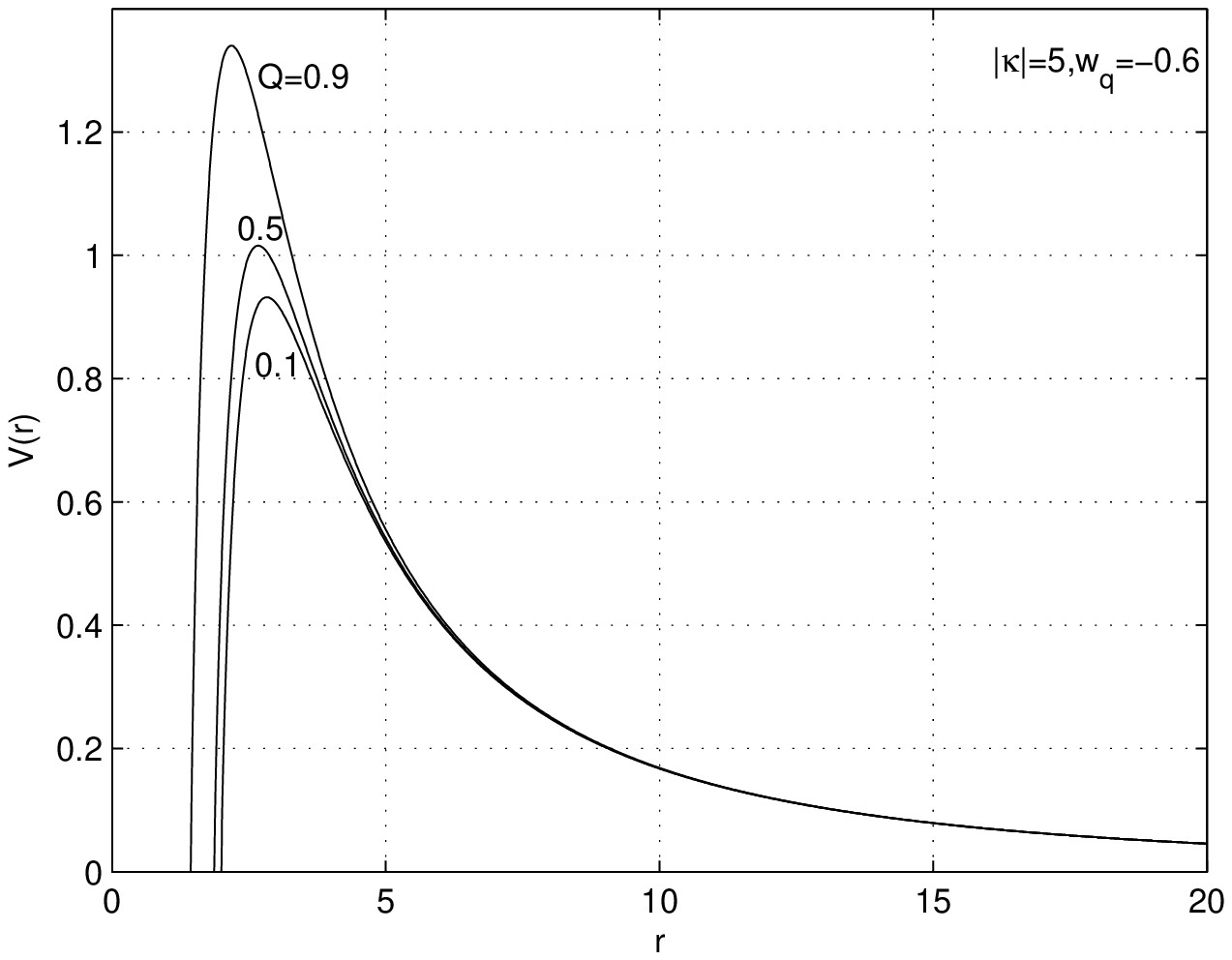}
\caption{Variation of the effective potential for the massless Dirac
field with $Q$ in RN black hole surrounded by quintessence for
$|\kappa|=5$, $w_{q}=-0.6$ and $c=0.001$.\label{fig:fig2}}
\end{figure}

\begin{figure}
    \includegraphics[angle=0, width=0.6\textwidth]{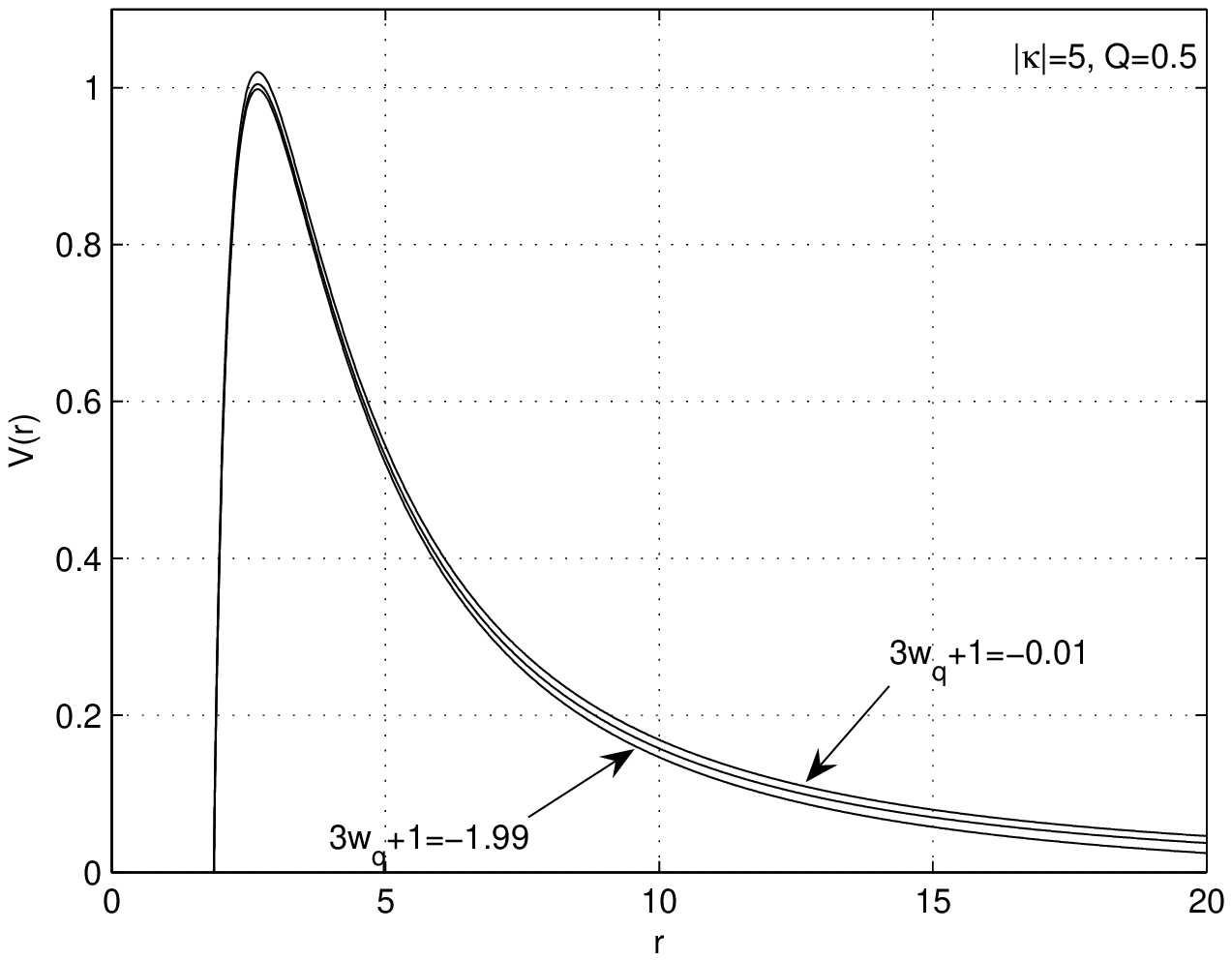}
\caption{Variation of the effective potential for the massless Dirac
field with $w_{q}$ in the RN black hole surrounded by quintessence
for $|\kappa|=5$, $Q=0.5$, $c=0.001$.\label{fig:fig3}}
\end{figure}

WKB method, a semianalytic technique for determining the quasinormal
mode complex frequencies of black holes was presented by Schutz and
Will\cite{Schutz} at the second order. Later, S. Iyer and C. M.
Will\cite{Iyer01} developed the method to the third order and R. A.
Konoplya\cite{Konoplya03} extended it to the sixth order. Generally
speaking, for the low-lying QNMs, the approach is quite accurately.
The method has been used to calculate the quasinormal frequencies
for the Schwarzschild\cite{Iyer02},
Reissner-Norstr\"{o}m\cite{Kokkotas02}, Kerr\cite{Seidel},
Kerr-Newman\cite{Kokkotas03} black holes, etc. The formula for the
complex quasinormal frequencies $\omega$(for the third order WKB
method) is given by
\begin{equation}\label{11}
    \omega^{2}=\left [V_{0}+(-2V''_{0})^{1/2}\Lambda\right ]-i(n+\frac{1}{2})(-2V''_{0})^{1/2}(1+\Omega)
\end{equation}
where
\begin{align*}
    \Lambda=\frac{1}{(-2V''_{0})^{1/2}}\left\{\frac{1}{8}\left(\frac{V^{(4)}_{0}}{V''_{0}}\right)(\frac{1}{4}+\alpha^{2})-\frac{1}{288}\left(\frac{V'''_{0}}{V''_{0}}\right)^{2}(7+60\alpha^{2})\right\}
\end{align*}
\begin{align}
   \Omega=&\,\frac{1}{-2V''_{0}}\Big\{\frac{5}{6912}\left(\frac{V'''_{0}}{V''_{0}}\right)^{4}(77+188\alpha^{2})\nonumber\\
   &-\frac{1}{384}\left(\frac{V'''^{2}_{0}V^{(4)}_{0}}{V''^{3}_{0}}\right)(51+100\alpha^{2})+\frac{1}{2304}\left(\frac{V^{(4)}_{0}}{V''_{0}}\right)^{2}(67+68\alpha^{2})\nonumber\\
    &+\frac{1}{288}\left(\frac{V'''_{0}V^{(5)}_{0}}{V''^{2}_{0}}\right)(19+28\alpha^{2})-\frac{1}{288}\left(\frac{V^{(6)}_{0}}{V''_{0}}\right )(5+4\alpha^{2})\Big \}\label{12}
\end{align}
and
\begin{equation}\label{13}
    \alpha=n+\frac{1}{2},\> V^{(n)}_{0}=\frac{d^{n}V}{dr^{n}_{*}}\Big|_{r_{*}=r_{*}(r_{p})}
\end{equation}
 We take $M=1, c=0.001$ and $M=1, c=0$ for our calculation. And $c=0$ means there is no quintessence.\\
Fig.\ref{fig:fig1} shows the variation of the effective potential
for the massless Dirac field with $|\kappa|$ in RN black hole
surrounded by quintessence for fixed $w_{q}=-0.6$, $Q=0.5$ and
$c=0.001$. From Fig.\ref{fig:fig1} we can find that the peak value
of potential barrier gets higher and the location of peak moves
along the right as $|\kappa|$ increases. Fig.\ref{fig:fig2} shows
the variation of the effective potential for the massless Dirac
field with $Q$ in RN black hole surrounded by quintessence for
$|\kappa|=5$, $w_{q}=-0.6$ and $c=0.001$. From Fig.\ref{fig:fig2} we
can see that as $Q$ increases the peak value of potential barrier
gets higher and the location of peak moves along the left.
Fig.\ref{fig:fig3} shows the variation of the effective potential
for the massless Dirac field with $w_{q}$ in the RN black hole
surrounded by quintessence for $|\kappa|=5$, $Q=0.5$ and $c=0.001$.
From Fig.\ref{fig:fig3} we can find that as the absolute value
$w_{q}$ increases the peak value of potential barrier gets
lower.\\
The quasinormal frequencies of the massless Dirac particles in the
Reissner-Nordstr\"{o}m  black hole surrounded by quintessence(for
fixed c=0.001) or no quintessence for different $Q$, $w_{q}$,
$|\kappa|$ and $n$ (for $0\leq n<\kappa|$)are presented in tables
\ref{tab:table1}-\ref{tab:table6} and figures
\ref{fig:fig4}-\ref{fig:fig7} by using the third WKB approximation.


 \setcounter{secnumdepth}{-1}

\section{conclusion and discussion}
\begin{figure}
    \includegraphics[angle=0, width=0.6\textwidth]{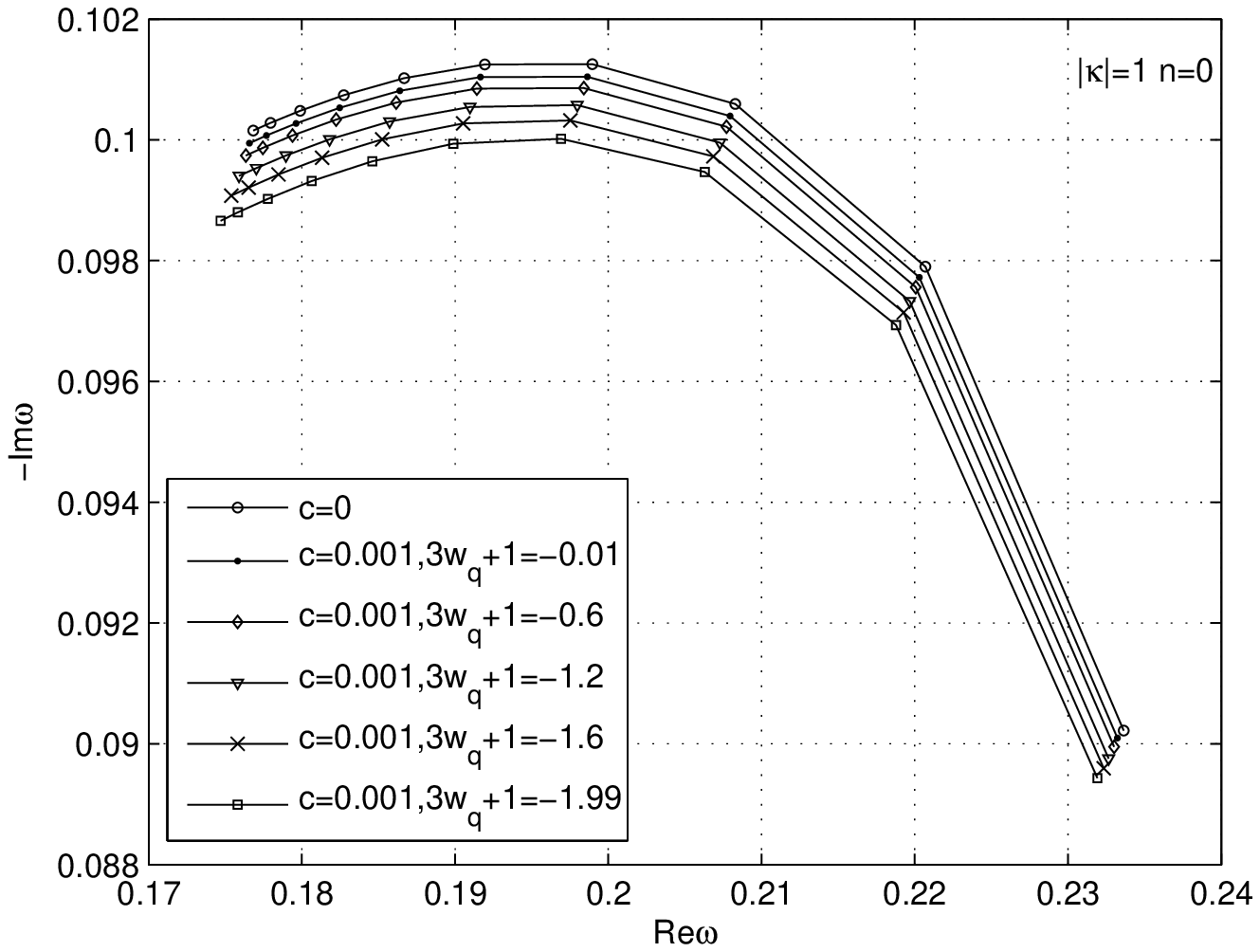}
    \includegraphics[angle=0, width=0.6\textwidth]{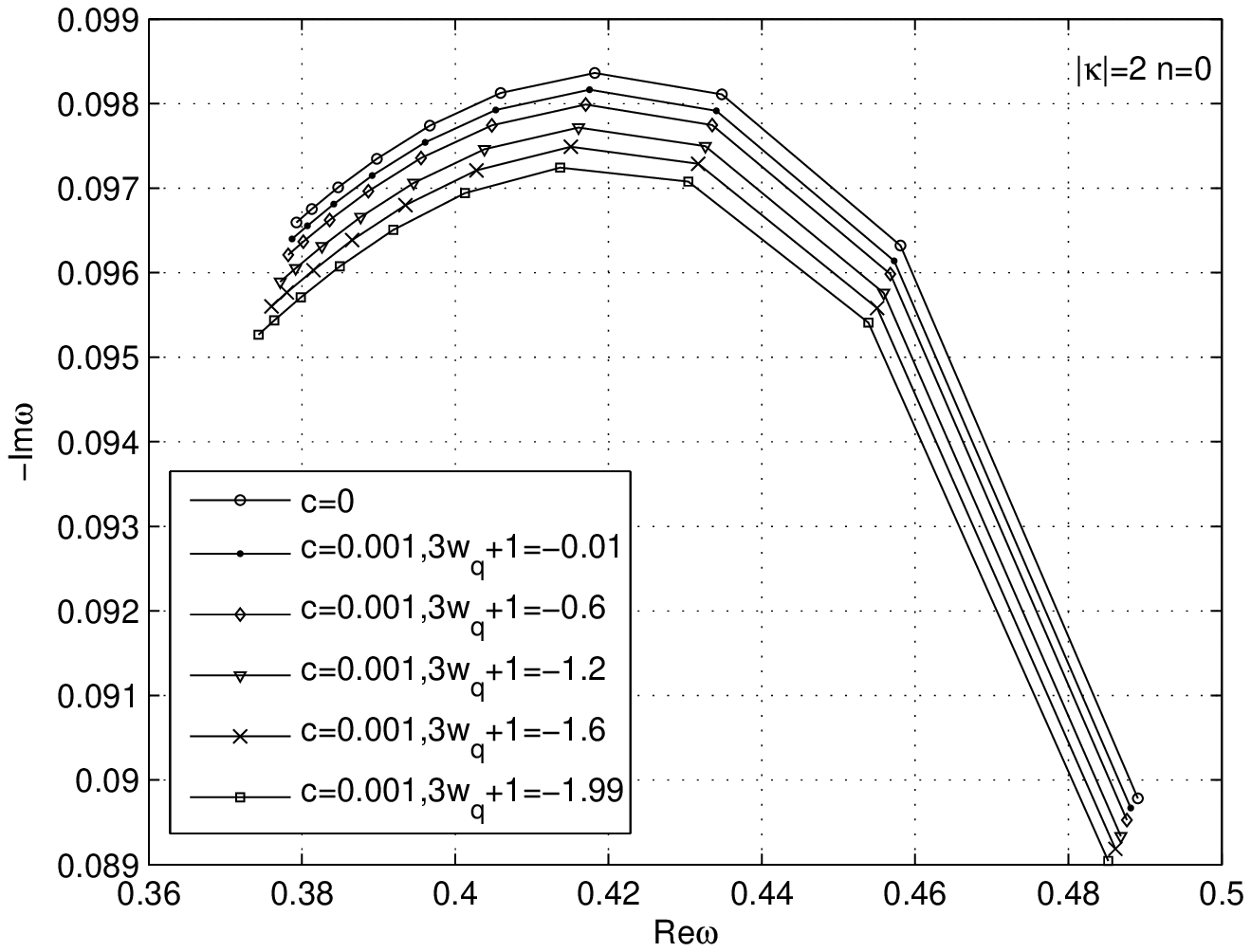}
    \includegraphics[angle=0, width=0.6\textwidth]{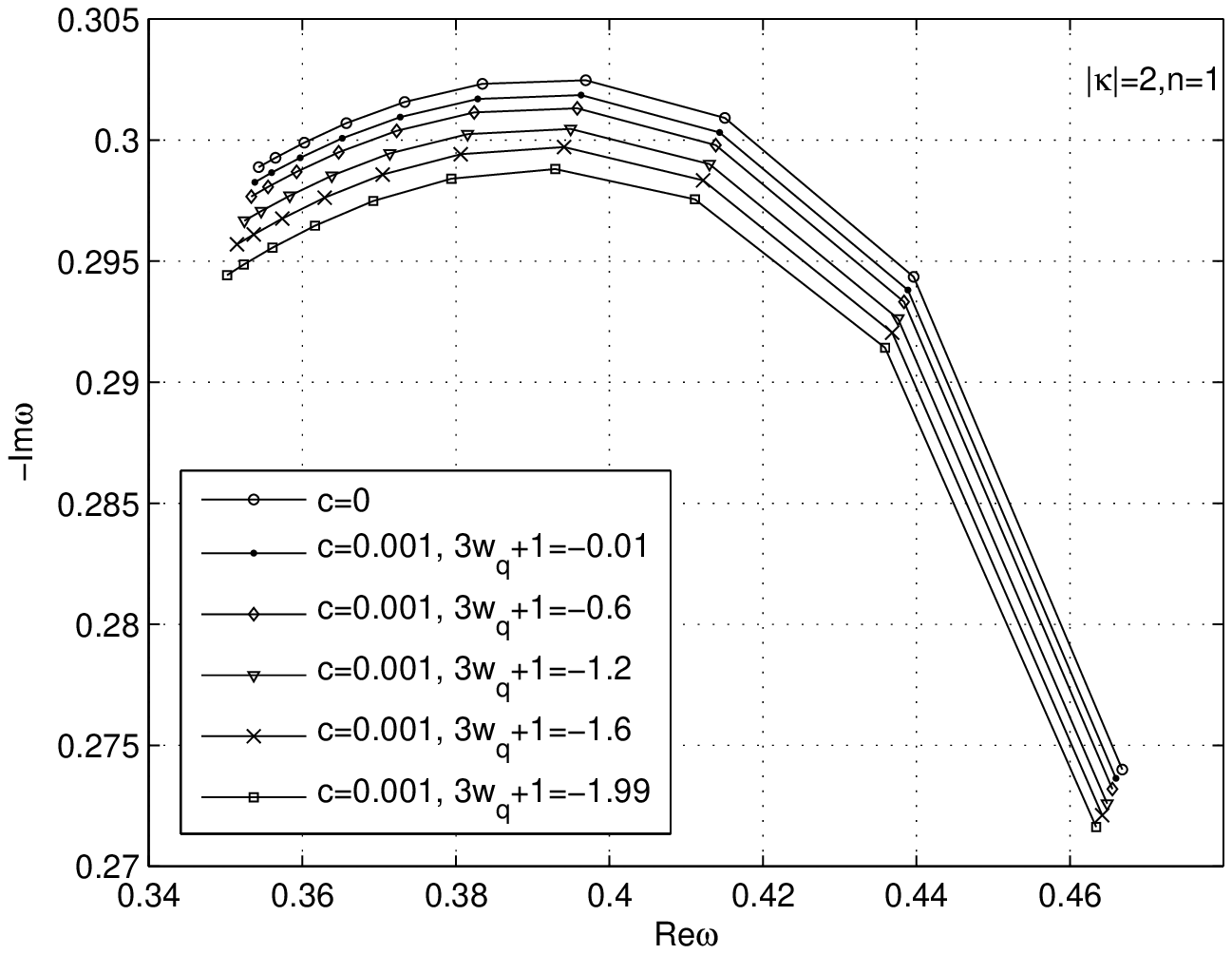}
\caption{The relationship between the real and imaginary
 parts of quasinormal frequencies for the massless Dirac
 perturbations in the background of the RN black hole surrounded by quintessence for
 fixed $c=0.001$ and no quintessence for $c=0$ with the different state
parameter $w_{q}$.\label{fig:fig4}}
\end{figure}

\begin{figure}
    \includegraphics[angle=0, width=0.6\textwidth]{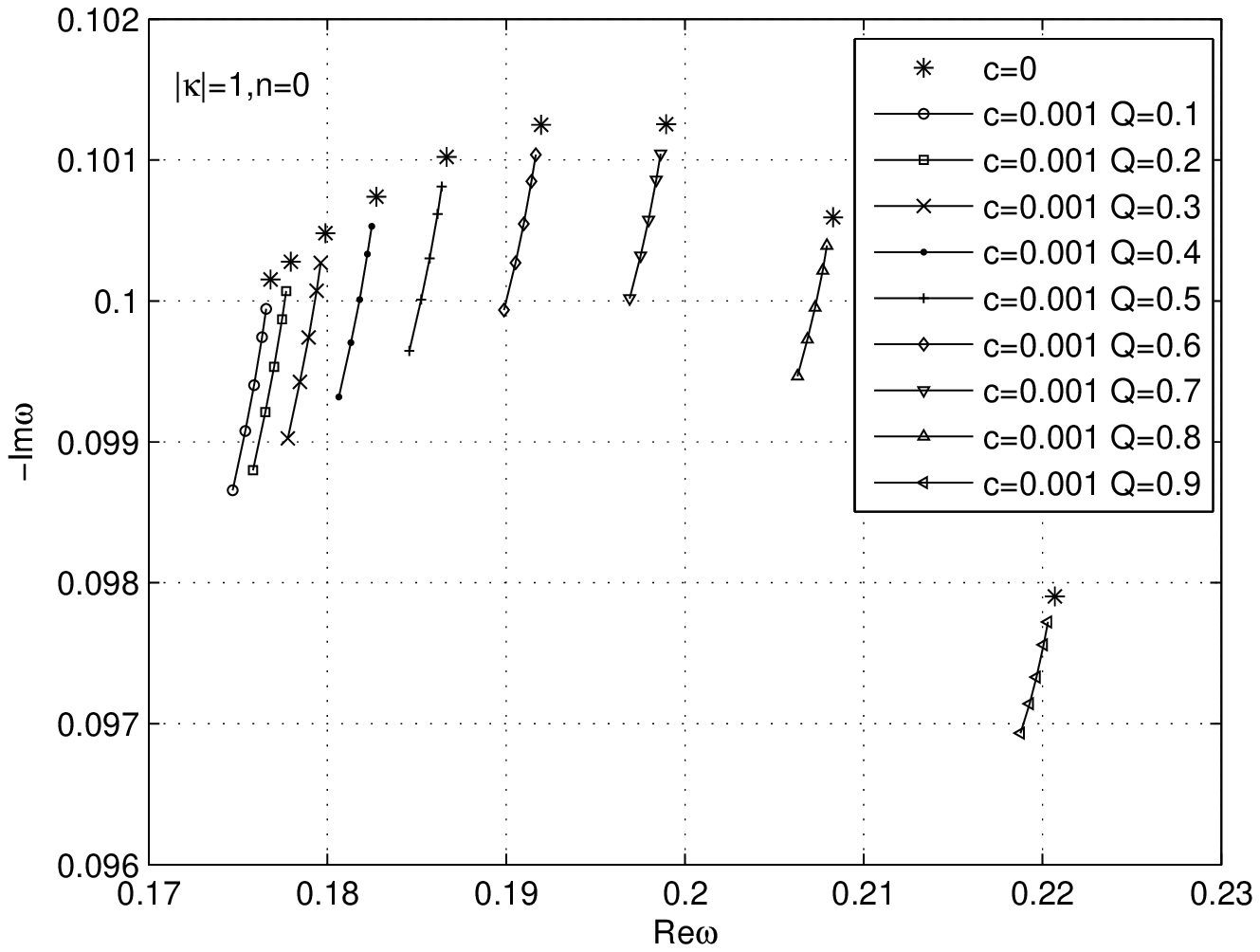}
    \includegraphics[angle=0, width=0.6\textwidth]{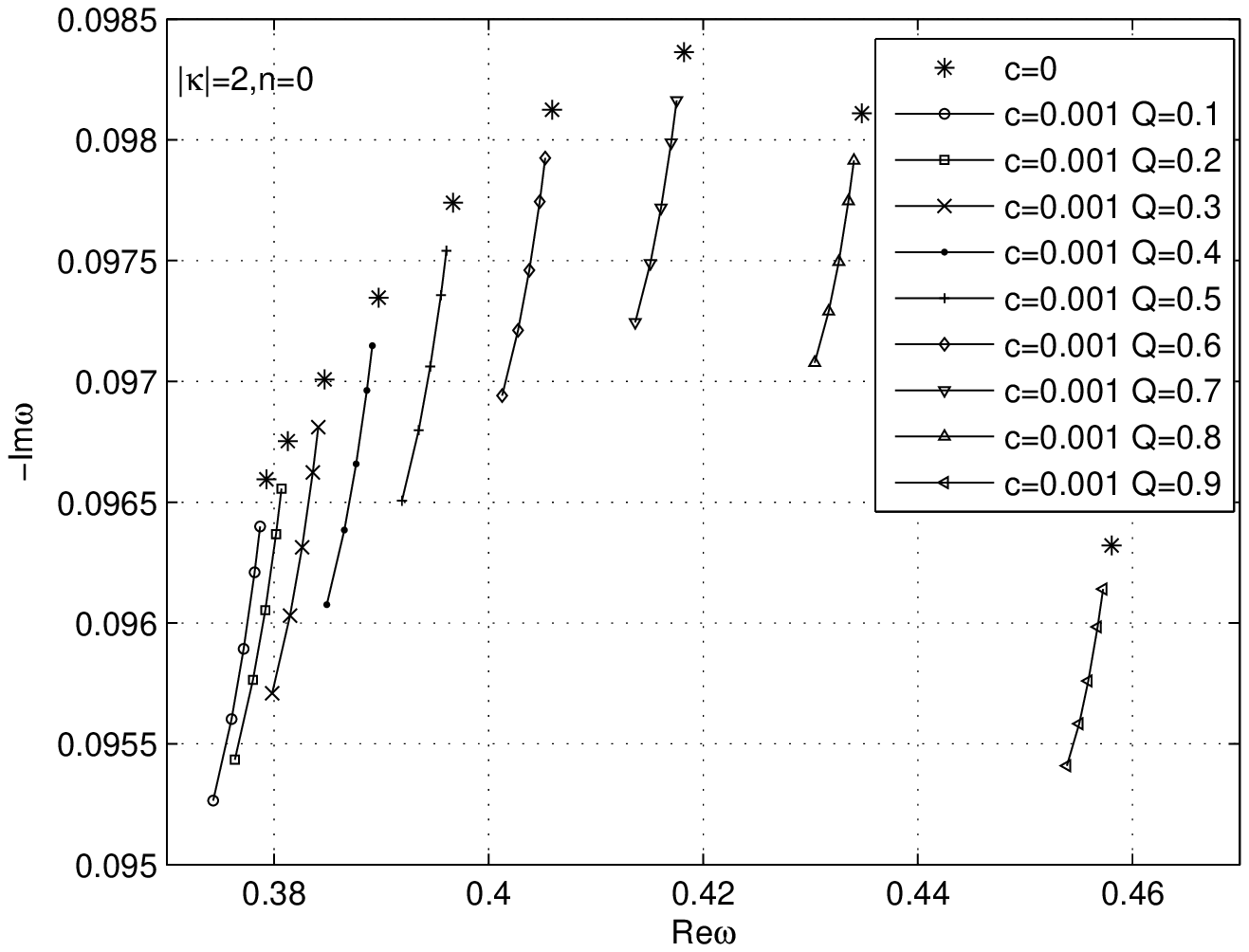}
    \includegraphics[angle=0, width=0.6\textwidth]{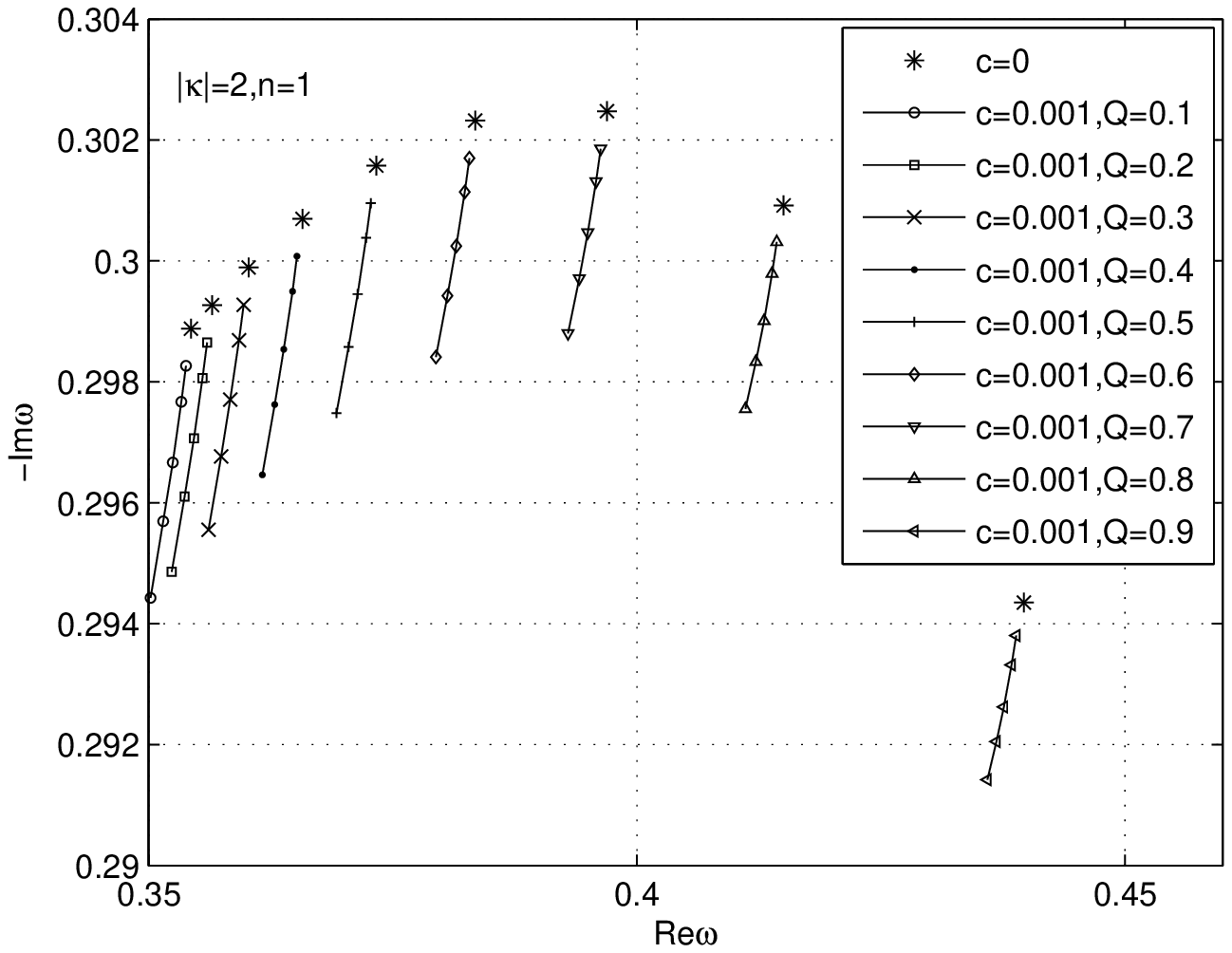}
\caption{The relationship between the real and imaginary
 parts of quasinormal frequencies for the massless Dirac
 perturbations in the background of the RN black hole surrounded by quintessence for
 fixed $c=0.001$ and no quintessence for $c=0$ with different $Q$.\label{fig:fig5}}
\end{figure}

\begin{figure}
    \includegraphics[angle=0, width=0.6\textwidth]{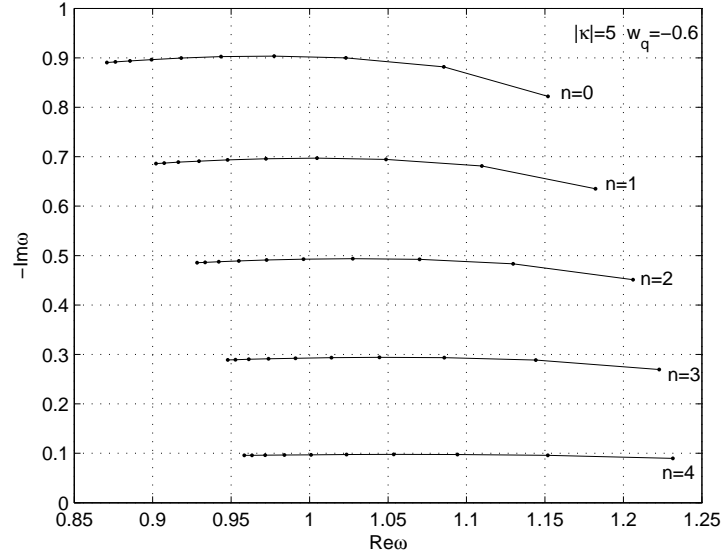}
\caption{The relationship between the real and imaginary
 parts of quasinormal frequencies for the massless Dirac
 perturbations in the background of the RN black hole surrounded by quintessence for $c=0.001$, $|\kappa|=5$, and $w_{q}=-0.6$ with different $n$.\label{fig:fig6}}
\end{figure}

\begin{figure}
    \includegraphics[angle=0, width=0.6\textwidth]{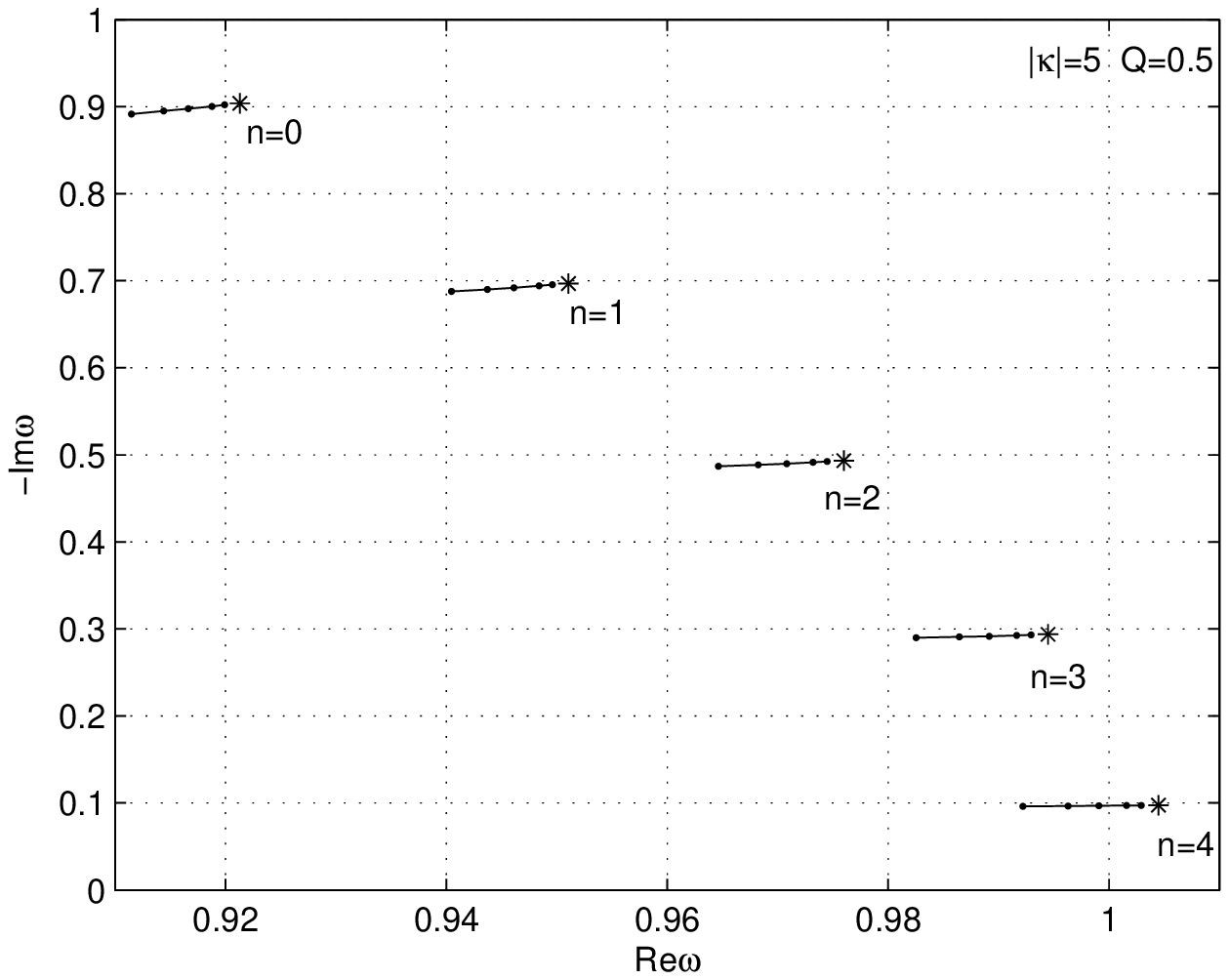}
\caption{The relationship between the real and imaginary
 parts of quasinormal frequencies for the massless Dirac
 perturbations in the background of the RN black hole surrounded by quintessence (for fixed $c=0.001$) and no quintessence ($c=0$) for $|\kappa|=5$ and $Q=0.5$ with different $n$.\label{fig:fig7}}
\end{figure}

We have calculated the low-lying QNM frequencies for massless Dirac
field in the Reissner-Nordstr\"{o}m black hole surrounded by
quintessence using the third WKB approximation. The results indicate
that the quasinormal frequencies not only depend on the mass M, the
charge Q, but also the quintessential state parameter $w_{q}$. \\
From the tables \ref{tab:table1}-\ref{tab:table6} and figures
\ref{fig:fig4}-\ref{fig:fig7} we find the main results as
follows:\\
a)The magnitude of the real part of the quasinormal mode frequencies
increases, while the magnitude of the imaginary one increases
firstly to a maximum at the vicinity of $Q = 0.7$ and then decreases
as the charge $Q$ increases for fixed $c=0.001$, $|\kappa|$, $n$ and
$w_{q}$. b)The magnitude of the real and the imaginary part of the
quasinormal mode frequencies decreases as the absolute value of
$w_{q}$ increases for fixed $c=0.001$, $|\kappa|$, $n$ and the
charge $Q$. It means that when the absolute value of $w_{q}$ is
larger, the oscillations damp more slowly. c)The magnitude of the
real and the imaginary part of quasinormal mode frequencies with
quintessence are smaller than those with no quintessence for fixed
$|\kappa|$ and $n$. Therefore, we conclude that due to the presence
of quintessence, the oscillations of the massless Dirac field damps
more slowly.

 \acknowledgements{This work was supported by the Research Programs of the
Educational Office of Liaoning Province of China under Grant No.
2009A036}

\begin{table}
\caption{\label{tab:table1} The quasinormal frequencies of massless
Dirac perturbations in the RN black hole without quintessence(c=0)
for $|\kappa|=1$, $|\kappa|=2$, $|\kappa|=5$}
         \begin{tabular}{ccccccc}
         \hline
      $|\kappa|$ & $Q$ & $\omega(n=0)$ & $|\kappa|$ & $Q$ & $\omega(n=0)$ & $\omega(n=1)$ \\
     \hline
     $$&$0.1$&$0.17682-0.10015i$&$$&$0.1$&$0.37928-0.09660i$&$0.35432-0.29888i$\\
     $$&$0.2$&$0.17795-0.10028i$&$$&$0.2$&$0.38128-0.09675i$&$0.35649-0.29927i$\\
     $$&$0.3$&$0.17989-0.10048i$&$$&$0.3$&$0.38472-0.09701i$&$0.36023-0.29989i$\\
     $$&$0.4$&$0.18274-0.10074i$&$$&$0.4$&$0.38976-0.09735i$&$0.36573-0.30070i$\\
     $$&$0.5$&$0.18668-0.10102i$&$$&$0.5$&$0.39668-0.09774i$&$0.37330-0.30158i$\\
     $1$&$0.6$&$0.19195-0.10125i$&$2$&$0.6$&$0.40593-0.09812i$&$0.38344-0.30232i$\\
     $$&$0.7$&$0.19895-0.10125i$&$$&$0.7$&$0.41821-0.09836i$&$0.39693-0.30247i$\\
     $$&$0.8$&$0.20829-0.10059i$&$$&$0.8$&$0.43479-0.09811i$&$0.41503-0.30092i$\\
     $$&$0.9$&$0.22069-0.09790i$&$$&$0.9$&$0.45808-0.09632i$&$0.43964-0.29435i$\\
     $$&$0.99$&$0.23364-0.09022i$&$$&$0.99$&$0.48907-0.08978i$&$0.46679-0.27401i$\\
     \hline
  \end{tabular}
      \begin{tabular}{ccccccc}
     \hline
     $|\kappa|$ & $Q$ & $\omega(n=0)$ & $\omega(n=1)$ & $\omega(n=2)$ & $\omega(n=3)$ &
     $\omega(n=4)$\\
     \hline
     $$&$0.1$&$0.96183-0.09631i$&$0.95123-0.29034i$&$0.93165-0.48789i$&$0.90530-0.68959i$&$0.87383-0.89485i$\\
     $$&$0.2$&$0.96674-0.09647i$&$0.95620-0.29080i$&$0.93674-0.48864i$&$0.91054-0.69061i$&$0.87926-0.89613i$\\
     $$&$0.3$&$0.97516-0.09672i$&$0.96474-0.29155i$&$0.94548-0.48986i$&$0.91955-0.69225i$&$0.88860-0.89818i$\\
     $$&$0.4$&$0.98753-0.09707i$&$0.97727-0.29256i$&$0.95832-0.49146i$&$0.93280-0.69441i$&$0.90235-0.90087i$\\
     $$&$0.5$&$1.00450-0.09747i$&$0.99449-0.29372i$&$0.97599-0.49330i$&$0.95106-0.69687i$&$0.92132-0.90390i$\\
     $5$&$0.6$&$1.02716-0.09786i$&$1.01751-0.29486i$&$0.99963-0.49507i$&$0.97552-0.69915i$&$0.94678-0.90664i$\\
     $$&$0.7$&$1.05731-0.09813i$&$1.04814-0.29558i$&$1.03113-0.49608i$&$1.00816-0.70027i$&$0.98076-0.90777i$\\
     $$&$0.8$&$1.09783-0.09857i$&$1.09018-0.29649i$&$1.07571-0.49643i$&$1.05563-0.69888i$&$1.03093-0.90369i$\\
     $$&$0.9$&$1.15585-0.09630i$&$1.14814-0.28974i$&$1.13360-0.48545i$&$1.11354-0.68405i$&$1.08918-0.88547i$\\
     $$&$0.99$&$1.23577-0.08994i$&$1.22708-0.27045i$&$1.21021-0.45273i$&$1.18603-0.63751i$&$1.15549-0.82508i$\\
    \hline
  \end{tabular}
\end{table}

\begin{table}
\caption{\label{tab:table2} Quasinormal frequencies of massless
Dirac perturbations in the RN black hole surrounded by quintessence
for $|\kappa|=1$, $|\kappa|=2$, $|\kappa|=5$, and $c=0.001$,
$3w_{q}+1=-0.01$}
         \begin{tabular}{ccccccc}
     \hline
      $|\kappa|$ & $Q$ & $\omega(n=0)$ & $|\kappa|$ & $Q$ & $\omega(n=0)$ & $\omega(n=1)$ \\
     \hline
     $$&$0.1$&$0.17657-0.09994i$&$$&$0.1$&$0.37872-0.09640i$&$0.35381-0.29826i$\\
     $$&$0.2$&$0.17770-0.10007i$&$$&$0.2$&$0.38071-0.09656i$&$0.35598-0.29865i$\\
     $$&$0.3$&$0.17963-0.10027i$&$$&$0.3$&$0.38414-0.09681i$&$0.35971-0.29928i$\\
     $$&$0.4$&$0.18248-0.10053i$&$$&$0.4$&$0.38917-0.09715i$&$0.36519-0.30008i$\\
     $$&$0.5$&$0.18640-0.10081i$&$$&$0.5$&$0.39607-0.09754i$&$0.37274-0.30096i$\\
     $1$&$0.6$&$0.19165-0.10104i$&$2$&$0.6$&$0.40529-0.09792i$&$0.38285-0.30170i$\\
     $$&$0.7$&$0.19864-0.10104i$&$$&$0.7$&$0.41754-0.09816i$&$0.39630-0.30186i$\\
     $$&$0.8$&$0.20795-0.10039i$&$$&$0.8$&$0.43406-0.09791i$&$0.41434-0.30031i$\\
     $$&$0.9$&$0.22030-0.09772i$&$$&$0.9$&$0.45727-0.09614i$&$0.43887-0.29380i$\\
     $$&$0.99$&$0.23323-0.09010i$&$$&$0.99$&$0.48813-0.08967i$&$0.46598-0.27364i$\\
     \hline
   \end{tabular}

       \begin{tabular}{ccccccc}
     \hline
     $|\kappa|$ & $Q$ & $\omega(n=0)$ & $\omega(n=1)$ & $\omega(n=2)$ & $\omega(n=3)$ &
     $\omega(n=4)$\\
     \hline
     $$&$0.1$&$0.96037-0.09611i$&$0.94980-0.28975i$&$0.93027-0.48689i$&$0.90398-0.68818i$&$0.87259-0.89301i$\\
     $$&$0.2$&$0.96527-0.09627i$&$0.95475-0.29021i$&$0.93534-0.48764i$&$0.90920-0.68920i$&$0.87800-0.89429i$\\
     $$&$0.3$&$0.97367-0.09653i$&$0.96327-0.29096i$&$0.94406-0.48886i$&$0.91819-0.69084i$&$0.88732-0.89634i$\\
     $$&$0.4$&$0.98600-0.09687i$&$0.97577-0.29196i$&$0.95687-0.49046i$&$0.93141-0.69299i$&$0.90103-0.89903i$\\
     $$&$0.5$&$1.00293-0.09727i$&$0.99295-0.29312i$&$0.97449-0.49230i$&$0.94962-0.69544i$&$0.91994-0.90204i$\\
     $5$&$0.6$&$1.02553-0.09766i$&$1.01590-0.29426i$&$0.99806-0.49406i$&$0.97401-0.69772i$&$0.94533-0.90478i$\\
     $$&$0.7$&$1.05558-0.09793i$&$1.04644-0.29498i$&$1.02947-0.49507i$&$1.00655-0.69884i$&$0.97921-0.90592i$\\
     $$&$0.8$&$1.09622-0.09774i$&$1.08775-0.29430i$&$1.07195-0.49361i$&$1.05052-0.69633i$&$1.02488-0.90219i$\\
     $$&$0.9$&$1.15379-0.09612i$&$1.14609-0.28920i$&$1.13158-0.48454i$&$1.11157-0.68277i$&$1.08726-0.88382i$\\
     $$&$0.99$&$1.23334-0.08983i$&$1.22470-0.27011i$&$1.20794-0.45214i$&$1.18390-0.63667i$&$1.15354-0.82395i$\\
    \hline
   \end{tabular}
 \end{table}

\begin{table}
\caption{\label{tab:table3} Quasinormal frequencies of massless
Dirac perturbations in the RN black hole surrounded by quintessence
for $|\kappa|=1$, $|\kappa|=2$, $|\kappa|=5$, and $c=0.001$,
$3w_{q}+1=-0.6$}
         \begin{tabular}{ccccccc}
     \hline
      $|\kappa|$ & $Q$ & $\omega(n=0)$ & $|\kappa|$ & $Q$ & $\omega(n=0)$ & $\omega(n=1)$ \\
     \hline
     $$&$0.1$&$0.17634-0.09974i$&$$&$0.1$&$0.37820-0.09621i$&$0.35334-0.29767i$\\
     $$&$0.2$&$0.17747-0.09987i$&$$&$0.2$&$0.38019-0.09637i$&$0.35551-0.29806i$\\
     $$&$0.3$&$0.17940-0.10007i$&$$&$0.3$&$0.38362-0.09662i$&$0.35924-0.29869i$\\
     $$&$0.4$&$0.18224-0.10033i$&$$&$0.4$&$0.38865-0.09696i$&$0.36472-0.29950i$\\
     $$&$0.5$&$0.18616-0.10062i$&$$&$0.5$&$0.39555-0.09736i$&$0.37227-0.30038i$\\
     $1$&$0.6$&$0.19142-0.10085i$&$2$&$0.6$&$0.40477-0.09774i$&$0.38237-0.30114i$\\
     $$&$0.7$&$0.19840-0.10086i$&$$&$0.7$&$0.41702-0.09799i$&$0.39582-0.30131i$\\
     $$&$0.8$&$0.20771-0.10022i$&$$&$0.8$&$0.43354-0.09775i$&$0.41385-0.29979i$\\
     $$&$0.9$&$0.22006-0.09756i$&$$&$0.9$&$0.45675-0.09598i$&$0.43839-0.29332i$\\
     $$&$0.99$&$0.23300-0.08995i$&$$&$0.99$&$0.48763-0.08953i$&$0.46553-0.27320i$\\
     \hline
   \end{tabular}
       \begin{tabular}{ccccccc}
     \hline
     $|\kappa|$ & $Q$ & $\omega(n=0)$ & $\omega(n=1)$ & $\omega(n=2)$ & $\omega(n=3)$ &
     $\omega(n=4)$\\
     \hline
     $$&$0.1$&$0.95904-0.09593i$&$0.94849-0.28918i$&$0.40477-0.09774i$&$0.90277-0.68682i$&$0.87144-0.89125i$\\
     $$&$0.2$&$0.96394-0.09608i$&$0.95345-0.28964i$&$0.93408-0.48669i$&$0.90799-0.68784i$&$0.87685-0.89254i$\\
     $$&$0.3$&$0.97234-0.09634i$&$0.96197-0.29040i$&$0.94280-0.48791i$&$0.91698-0.68949i$&$0.88616-0.89460i$\\
     $$&$0.4$&$0.98467-0.09668i$&$0.97447-0.29140i$&$0.95560-0.48952i$&$0.93020-0.69166i$&$0.89987-0.89730i$\\
     $$&$0.5$&$1.00160-0.09708i$&$0.99164-0.29257i$&$0.97322-0.49137i$&$0.94839-0.69412i$&$0.91878-0.90034i$\\
     $5$&$0.6$&$1.02420-0.09749i$&$1.01459-0.29372i$&$0.99679-0.49315i$&$0.97278-0.69643i$&$0.94415-0.90311i$\\
     $$&$0.7$&$1.05426-0.09776i$&$1.04513-0.29446i$&$1.02820-0.49418i$&$1.00532-0.69759i$&$0.97802-0.90429i$\\
     $$&$0.8$&$1.09490-0.09757i$&$1.08644-0.29379i$&$1.07068-0.49276i$&$1.04929-0.69513i$&$1.02369-0.90063i$\\
     $$&$0.9$&$1.15248-0.09596i$&$1.14480-0.28873i$&$1.13032-0.48374i$&$1.11034-0.68165i$&$1.08608-0.88236i$\\
     $$&$0.99$&$1.23207-0.08968i$&$1.22344-0.26968i$&$1.20671-0.45143i$&$1.18273-0.63565i$&$1.15243-0.82264i$\\
    \hline
   \end{tabular}
 \end{table}

\begin{table}
\caption{\label{tab:table4} Quasinormal frequencies of massless
Dirac perturbations in the RN black hole surrounded by quintessence
for $|\kappa|=1$, $|\kappa|=2$, $|\kappa|=5$, and $c=0.001$,
$3w_{q}+1=-1.2$}
         \begin{tabular}{ccccccc}
     \hline
      $|\kappa|$ & $Q$ & $\omega(n=0)$ & $|\kappa|$ & $Q$ & $\omega(n=0)$ & $\omega(n=1)$ \\
     \hline
     $$&$0.1$&$0.17590-0.09940i$&$$&$0.1$&$0.37719-0.09589i$&$0.35245-0.29667i$\\
     $$&$0.2$&$0.17702-0.09953i$&$$&$0.2$&$0.37919-0.09605i$&$0.35462-0.29707i$\\
     $$&$0.3$&$0.17896-0.09974i$&$$&$0.3$&$0.38262-0.09631i$&$0.35835-0.29771i$\\
     $$&$0.4$&$0.18180-0.10001i$&$$&$0.4$&$0.38766-0.09666i$&$0.36384-0.29854i$\\
     $$&$0.5$&$0.18572-0.10030i$&$$&$0.5$&$0.39457-0.09706i$&$0.37139-0.29945i$\\
     $1$&$0.6$&$0.19098-0.10055i$&$2$&$0.6$&$0.40381-0.09746i$&$0.38150-0.30025i$\\
     $$&$0.7$&$0.19797-0.10057i$&$$&$0.7$&$0.41608-0.09772i$&$0.39496-0.30047i$\\
     $$&$0.8$&$0.20728-0.09995i$&$$&$0.8$&$0.43264-0.09750i$&$0.41302-0.29901i$\\
     $$&$0.9$&$0.21966-0.09733i$&$$&$0.9$&$0.45590-0.09576i$&$0.43760-0.29262i$\\
     $$&$0.99$&$0.23265-0.08975i$&$$&$0.99$&$0.48685-0.08933i$&$0.46483-0.27259i$\\
     \hline
   \end{tabular}
       \begin{tabular}{ccccccc}
     \hline
     $|\kappa|$ & $Q$ & $\omega(n=0)$ & $\omega(n=1)$ & $\omega(n=2)$ & $\omega(n=3)$ &
     $\omega(n=4)$\\
     \hline
     $$&$0.1$&$0.95644-0.09562i$&$0.94595-0.28824i$&$0.92657-0.48433i$&$0.90047-0.68453i$&$0.86928-0.88824i$\\
     $$&$0.2$&$0.96135-0.09577i$&$0.95091-0.28870i$&$0.93163-0.48509i$&$0.90567-0.68557i$&$0.87467-0.88957i$\\
     $$&$0.3$&$0.96977-0.09603i$&$0.95945-0.28947i$&$0.94037-0.48633i$&$0.91467-0.68725i$&$0.88399-0.89167i$\\
     $$&$0.4$&$0.98213-0.09638i$&$0.97197-0.29049i$&$0.95319-0.48797i$&$0.92790-0.68946i$&$0.89771-0.89443i$\\
     $$&$0.5$&$0.99909-0.09679i$&$0.98918-0.29168i$&$0.97084-0.48987i$&$0.94612-0.69199i$&$0.91663-0.89755i$\\
     $5$&$0.6$&$1.02174-0.09720i$&$1.01217-0.29286i$&$0.99445-0.49171i$&$0.97054-0.69438i$&$0.94202-0.90043i$\\
     $$&$0.7$&$1.05186-0.09749i$&$1.04277-0.29364i$&$1.02590-0.49281i$&$1.00311-0.69564i$&$0.97592-0.90176i$\\
     $$&$0.8$&$1.09259-0.09733i$&$1.08416-0.29304i$&$1.06846-0.49149i$&$1.04715-0.69332i$&$1.02164-0.89828i$\\
     $$&$0.9$&$1.15030-0.09573i$&$1.14265-0.28805i$&$1.12822-0.48261i$&$1.10831-0.68004i$&$1.08414-0.88027i$\\
     $$&$0.99$&$1.23008-0.08949i$&$1.22149-0.26909i$&$1.20481-0.45044i$&$1.18091-0.63425i$&$1.15071-0.82082i$\\
     \hline
   \end{tabular}
 \end{table}

\begin{table}
\caption{\label{tab:table5}Quasinormal frequencies of massless Dirac
perturbations in the RN black hole surrounded by quintessence for
$|\kappa|=1$, $|\kappa|=2$, $|\kappa|=5$, and $c=0.001$,
$3w_{q}+1=-1.6$}
         \begin{tabular}{ccccccc}
     \hline
      $|\kappa|$ & $Q$ & $\omega(n=0)$ & $|\kappa|$ & $Q$ & $\omega(n=0)$ & $\omega(n=1)$ \\
     \hline
     $$&$0.1$&$0.17541-0.09907i$&$$&$0.1$&$0.37604-0.09560i$&$0.35149-0.29569i$\\
     $$&$0.2$&$0.17653-0.09921i$&$$&$0.2$&$0.37805-0.09577i$&$0.35366-0.29610i$\\
     $$&$0.3$&$0.17847-0.09943i$&$$&$0.3$&$0.38149-0.09603i$&$0.35740-0.29677i$\\
     $$&$0.4$&$0.18132-0.09970i$&$$&$0.4$&$0.38655-0.09638i$&$0.36290-0.29762i$\\
     $$&$0.5$&$0.18525-0.10001i$&$$&$0.5$&$0.39349-0.09680i$&$0.37047-0.29858i$\\
     $1$&$0.6$&$0.19052-0.10027i$&$2$&$0.6$&$0.40276-0.09721i$&$0.38060-0.29942i$\\
     $$&$0.7$&$0.19752-0.10032i$&$$&$0.7$&$0.41508-0.09749i$&$0.39409-0.29971i$\\
     $$&$0.8$&$0.20686-0.09973i$&$$&$0.8$&$0.43170-0.09729i$&$0.41219-0.29834i$\\
     $$&$0.9$&$0.21928-0.09714i$&$$&$0.9$&$0.45504-0.09558i$&$0.43684-0.29205i$\\
     $$&$0.99$&$0.23233-0.08960i$&$$&$0.99$&$0.48612-0.08919i$&$0.46418-0.27212i$\\
     \hline
     \end{tabular}
         \begin{tabular}{ccccccc}
     \hline
     $|\kappa|$ & $Q$ & $\omega(n=0)$ & $\omega(n=1)$ & $\omega(n=2)$ & $\omega(n=3)$ &
     $\omega(n=4)$\\
     \hline
     $$&$0.1$&$0.95347-0.09533i$&$0.94306-0.28735i$&$0.92383-0.48281i$&$0.89791-0.68233i$&$0.86694-0.88532i$\\
     $$&$0.2$&$0.95839-0.09549i$&$0.94805-0.28783i$&$0.92892-0.48360i$&$0.90316-0.68339i$&$0.87237-0.88667i$\\
     $$&$0.3$&$0.96685-0.09576i$&$0.95662-0.28862i$&$0.93769-0.48487i$&$0.91218-0.68512i$&$0.88171-0.88883i$\\
     $$&$0.4$&$0.97926-0.09611i$&$0.96919-0.28966i$&$0.95056-0.48655i$&$0.92545-0.68740i$&$0.89545-0.89168i$\\
     $$&$0.5$&$0.99630-0.09653i$&$0.98646-0.29089i$&$0.96826-0.48851i$&$0.94371-0.69001i$&$0.91441-0.89491i$\\
     $5$&$0.6$&$1.01905-0.09696i$&$1.00955-0.29212i$&$0.99195-0.49043i$&$0.96820-0.69252i$&$0.93985-0.89795i$\\
     $$&$0.7$&$1.04929-0.09726i$&$1.04026-0.29295i$&$1.02351-0.49163i$&$1.00086-0.69392i$&$0.97382-0.89946i$\\
     $$&$0.8$&$1.09018-0.09712i$&$1.08181-0.29242i$&$1.06620-0.49042i$&$1.04501-0.69178i$&$1.01964-0.89623i$\\
     $$&$0.9$&$1.14811-0.09556i$&$1.14050-0.28752i$&$1.12615-0.48170i$&$1.10635-0.67873i$&$1.08230-0.87852i$\\
     $$&$0.99$&$1.22819-0.08934i$&$1.21964-0.26865i$&$1.20303-0.44968i$&$1.17923-0.63316i$&$1.14914-0.81938i$\\
    \hline
    \end{tabular}
 \end{table}

\begin{table}
\caption{\label{tab:table6}Quasinormal frequencies of massless Dirac
perturbations in the RN black hole surrounded by quintessence for
$|\kappa|=1$, $|\kappa|=2$, $|\kappa|=5$, and $c=0.001$,
$3w_{q}+1=-1.99$}
         \begin{tabular}{ccccccc}
     \hline
      $|\kappa|$ & $Q$ & $\omega(n=0)$ & $|\kappa|$ & $Q$ & $\omega(n=0)$ & $\omega(n=1)$ \\
     \hline
     $$&$0.1$&$0.17470-0.09866i$&$$&$0.1$&$0.37434-0.09527i$&$0.35017-0.29443i$\\
     $$&$0.2$&$0.17584-0.09880i$&$$&$0.2$&$0.37637-0.09543i$&$0.35235-0.29486i$\\
     $$&$0.3$&$0.17778-0.09902i$&$$&$0.3$&$0.37984-0.09571i$&$0.35611-0.29555i$\\
     $$&$0.4$&$0.18065-0.09932i$&$$&$0.4$&$0.38494-0.09608i$&$0.36163-0.29646i$\\
     $$&$0.5$&$0.18459-0.09965i$&$$&$0.5$&$0.39193-0.09651i$&$0.36923-0.29748i$\\
     $1$&$0.6$&$0.18988-0.09993i$&$2$&$0.6$&$0.40127-0.09694i$&$0.37941-0.29841i$\\
     $$&$0.7$&$0.19691-0.10002i$&$$&$0.7$&$0.41367-0.09724i$&$0.39295-0.29880i$\\
     $$&$0.8$&$0.20629-0.09947i$&$$&$0.8$&$0.43041-0.09708i$&$0.41113-0.29755i$\\
     $$&$0.9$&$0.21878-0.09693i$&$$&$0..9$&$0.45390-0.09541i$&$0.43589-0.29142i$\\
     $$&$0.99$&$0.23192-0.08943i$&$$&$0.99$&$0.48518-0.08904i$&$0.46340-0.27163i$\\
     \hline
    \end{tabular}
        \begin{tabular}{ccccccc}
     \hline
     $|\kappa|$ & $Q$ & $\omega(n=0)$ & $\omega(n=1)$ & $\omega(n=2)$ & $\omega(n=3)$ &
     $\omega(n=4)$\\
     \hline
     $$&$0.1$&$0.94899-0.09501i$&$0.93879-0.28635i$&$0.91992-0.48099i$&$0.89442-0.67953i$&$0.86386-0.88142i$\\
     $$&$0.2$&$0.95396-0.09518i$&$0.94382-0.28684i$&$0.92505-0.48180i$&$0.89969-0.68064i$&$0.86930-0.88282i$\\
     $$&$0.3$&$0.96250-0.09545i$&$0.95246-0.28766i$&$0.93387-0.48312i$&$0.90876-0.68244i$&$0.87868-0.88509i$\\
     $$&$0.4$&$0.97502-0.09582i$&$0.96514-0.28874i$&$0.94683-0.48488i$&$0.92210-0.68483i$&$0.89248-0.88809i$\\
     $$&$0.5$&$0.99220-0.09626i$&$0.98254-0.29002i$&$0.96465-0.48693i$&$0.94046-0.68758i$&$0.91151-0.89152i$\\
     $5$&$0.6$&$1.01513-0.09670i$&$1.00580-0.29131i$&$0.98848-0.48896i$&$0.96507-0.69027i$&$0.93705-0.89481i$\\
     $$&$0.7$&$1.04561-0.09703i$&$1.03673-0.29223i$&$1.02023-0.49031i$&$0.99789-0.69190i$&$0.97114-0.89663i$\\
     $$&$0.8$&$1.08681-0.09692i$&$1.07856-0.29179i$&$1.06318-0.48928i$&$1.04225-0.69004i$&$1.01714-0.89378i$\\
     $$&$0.9$&$1.14514-0.09539i$&$1.13764-0.28701i$&$1.12346-0.48077i$&$1.10388-0.67732i$&$1.08005-0.87655i$\\
     $$&$0.99$&$1.22576-0.08920i$&$1.21727-0.26822i$&$1.20080-0.44894i$&$1.17716-0.63206i$&$1.14727-0.81787i$\\
     \hline
    \end{tabular}
 \end{table}

\end{document}